\documentclass[letterpaper,twocolumn,10pt]{article}
\usepackage{usenix2019_v3}

\usepackage{tikz}
\usepackage{amsmath}
\usepackage{authblk}
\usepackage[english]{babel}
\usepackage[linesnumbered,ruled,lined]{algorithm2e}
\usepackage{subfigure}
\usepackage{graphicx}
\usepackage{array}
\usepackage{pifont}
\usepackage{algpseudocode}
\usepackage{enumitem}

\usepackage{filecontents}

\title{Exploring the Initial Performance of NB-IoT NTN over GEO: Measurement and Analysis}

\begin{document}

\date{}


\author[1]{\textrm{Xiong Wang}}
\author[2]{\textrm{Yumin Du}}
\author[3]{\textrm{Bin Gu}}
\author[4]{\textrm{Zheng Lin}}
\author[2]{\textrm{Xiaoyang Li}}
\author[2]{\textrm{Yi Gong}}
\author[1]{\textrm{Wei Gong}}
\author[5]{\textrm{Linghe Kong}}

\affil[1]{University of Science and Technology of China}
\affil[2]{Southern University of Science and Technology}
\affil[3]{Independent Researcher} 
\affil[4]{University of Luxembourg}
\affil[5]{Shanghai Jiao Tong University, China}


\maketitle

\begin{abstract}
With the standardization of Non-Terrestrial Networks (NTN) to provide direct satellite connectivity to massive, low-power Internet of Things (IoT) devices, 3GPP IoT-NTN bridges the worlds of cellular and satellite communications. While holding great potential for global connectivity with IoT devices, there exist several concerns about the system performance of IoT-NTN over Geostationary Earth Orbit (GEO), which covers multiple dimensions such as end-to-end delay and energy consumption considering the ultra-long transmission distance from ground IoT terminals to GEO satellite. To answer these concerns, we have conducted the first comprehensive and in-depth measurement for NB-IoT NTN over GEO. Based on real NB-IoT NTN testbeds including both Skylo and Tiantong, measurements covering more than six months confirm that the available implementation of NB-IoT NTN remains in the initial stage. Amplified by ultra-long Round-Trip Time (RTT) between ground IoT terminals and GEO satellite, there exists plenty of time and energy consumption during the access process. Interestingly, it also reveals that as an energy-saving mechanism, Power Saving Mode (PSM) fundamentally reshapes NB-IoT NTN traffic into a bursty and access-driven communication pattern and thus has a considerable influence on the end-to-end delay and energy consumption. Finally, we propose corresponding optimization schemes to reduce the delay and energy, which lays a good foundation for the implementation of NB-IoT NTN in the near future.    
\end{abstract}

\section{Introduction}
Coupled with a rapidly expanding segment of global IoT deployments~\cite{lin2026gapsl,fang2026hfedmoe,sun2025rrto,zhang2025robust,lin2025hasfl,yuan2023graph,fang2024automated} (such as maritime monitoring, environmental sensing, agriculture, energy infrastructure, logistics tracking, and emergency communications.) that cannot be economically served by terrestrial networks alone \cite{wang2025performance,lin2024fedsn,zhang2024satfed,teng2026communication,yuan2024satsense,peng2025sigchord,lin2025leo}, the Third Generation Partnership Project (3GPP) has progressively incorporated Non-Terrestrial Networks (NTN) into its standardization roadmap to extend cellular IoT services beyond terrestrial coverage. Initially, official support has been specified in Release 17 by introducing NB-IoT and LTE-M for satellite links, primarily targeting Low Earth Orbit (LEO) and Geostationary Earth Orbit (GEO) systems \cite{maattanen20243gpp,lin2025bridge}.

Meanwhile, industry prediction highlights that the overall satellite-IoT market will reach several tens of billions by 2030, as well as satellite-enabled IoT connections will hit hundreds of millions by the early 2030s \cite{paolini2025strategic}. Within this ecosystem, 3GPP based IoT-NTN is expected to capture a substantial share due to its strong standardization, interoperability, and integration with existing cellular infrastructure.

As an specific user case, NB-IoT NTN over the mature GEO system has been introduced to connect low-power Internet of Things (IoT) devices globally deployed due to NB-IoT's outstanding anti-noise capability \cite{giambene2024design,plastras2024non}. While suitable for terrestrial networks under short RTT and relatively stable links, the real-world behavior of NB-IoT based NTN featured with ultra-long RTT and high dynamic links remains unexplored and leaves a significant gap between standard specifications and practical performance, especially to developers and academic researchers. However, investigating the real-world performance of NB-IoT NTN over GEO is a non-trivial task due to the following challenges. First, NB-IoT is a proprietary wireless technique operating on licensed spectrum, and scare academic IoT-NTN testbed makes real-world measurements more challenging. Secondly, considering the infant stage with few IoT-NTN applications, performing IoT-NTN measurements is much more challenging compared to the mature terrestrial networks such as cellular networks \cite{hassan2022vivisecting,xu2020understanding} and terrestrial NB-IoT networks \cite{yang2020understanding,martinez2019exploring}.

\begin{figure}[t]
  	\centering
  	\includegraphics[width=6cm]{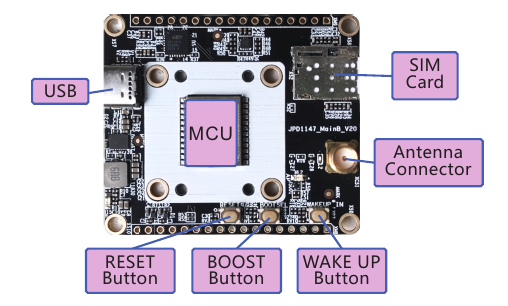}
	\caption{JMD 47 mainboard for NB-IoT NTN field test.}
    \label{hardware}
\end{figure}

Therefore, we present the first end-to-end measurement study of commercial NB-IoT NTN based on two well-known satellite service providers--Skylo \cite{saarnisaari2024military} and Tiantong \cite{zheng2021design}, revealing fundamental mismatches between the standard design and real satellite deployments. By deploying the device--JMD 47 \cite{JMD} as ground terminals and is shown in Fig.\ref{hardware}, measurements lasting for more than 6 months mainly cover two critical metrics in NB-IoT NTN--End-To-End (E2E) delay and energy consumption. Measurement results uncover that there exists a linear relationship between E2E delay and energy consumption, and in addition to concurrency level, both payload size and traffic interval affect access competition and E2E delay due to ultra-long RTT. 


\textbf{Measurement perspectives.} This initial measurement targets at investigating the critical performance metrics in NB-IoT NTN over GEO and cover the points as below. 

\textit{(i) E2E delay under different payload size.} Considering the ultra-long distance (around 36000 km) between the ground terminal and GEO satellite, the generated long transmission delay around 500 ms and weak reception power amplify the interaction between the ground terminal and GEO satellite. Hence, we measured the E2E delay under different payload size of packets.

\textit{(ii) E2E delay under different traffic interval.} Meanwhile, considering the ultra-long transmission delay around 500 ms, we have evaluated the performance of E2E delay under different traffic interval, which can achieved by setting 2, 5, and 10 seconds between consecutively transmitted packets. 


\textit{(iii) E2E delay under different concurrency level.} To evaluate NB-IoT NTN in concurrent scenarios, we have deployed ten ground terminals to transmit packets to the GEO satellite simultaneously. Compared to factors such as payload size and traffic interval that are negligible in terrestrial networks, the concurrency level has a significant impact on E2E delay in both terrestrial and Non-Terrestrial Networks (NTN). 

\textit{(iv) Energy VS delay.} As another critical metric in NB-IoT NTN over GEO, we first investigate the relationship between the energy consumption and E2E delay. Meanwhile, we have obtained the power consumption trace of the ground terminal by using the UT70 USB measurement device \cite{glavavs2024analysis}, and align them with the operations specified in the access process of NB-IoT NTN.

\textbf{Summary of insights.} Our large-scale measurements contribute to several major insights, as listed below.

\textit{(i) It reveals that the payload size of packets has a non-negligible influence on the E2E delay due to the limited Transport Block Size (TBS) specified in NB-IoT NTN uplink transmissions. E2E delay is control-plane dominated rather than traffic-dependent. However, this factor is almost negligible with regard to the E2E delay in terrestrial networks.} 

\textit{(ii) As a more mature NB-IoT NTN service provider, more suitable settings such as small TBS configuration are observed in Skylo compared to Tiantong, which implies that parameters derived from analytical assumptions and laboratory emulation remain far from functioning well in actual NB-IoT NTN. It also implies that NB-IoT NTN over GEO is small-packet optimized system, which is consistent with the observation that small packets can be delivered coupled with MSG3 or NAS due to the ultra-long RTT.}

\textit{(iii) Setting different traffic interval also affects the E2E delay performance due to MAC access and HARQ/NPUSCH retries, which is intensified by the short interval between packets due to the ultra-long RTT between the ground terminal and GEO satellite. In concurrent scenarios, access delay increases linearly with the number of concurrently transmitted ground terminals, which incurs long-tail latency problem.} 

\textit{(iv) In NB-IoT NTN over GEO, the energy consumption increases linearly with E2E delay, and it is mainly incurred by retransmission of MSG3 (i.e., retransmission number for packet reception) and cumbersome NAS overhead, where data transmission only occupies a small portion of 5\%. Thus, cellular protocol design fundamentally mismatches GEO satellite environments. Meanwhile, as a power saving mechanism, Power Saving Mode (PSM) fundamentally reshapes NB-IoT NTN traffic into a bursty and access-driven communication pattern, which should be carefully aligned with the traffic pattern.}

The contributions of this paper are summarized as below.
\begin{itemize}[topsep=0pt,itemsep=0pt,parsep=\parskip]
    \item We present the first comprehensive and in-depth measurement and investigation based on two commercial NB-IoT NTN service providers--Skylo and Tiantong, which mainly covers two critical aspects such as end-to-end delay and energy consumption.
    \item  We model both the end-to-end delay and energy consumption and introduce a breakdown analysis which pinpoints the bottleneck and space for improvement.
    \item We reveal that PSM transforms delay from a channel-dominated metric into an access- and traffic-driven phenomenon, where packet interval and payload size jointly determine the delay.
   \item To mediate the inefficiencies within the existing NB-IoT NTN system, which are derived from the above measurement and analysis, we propose several simple yet effective designs and conduct trace-driven simulations to verify their high efficiency. 
    
\end{itemize}
\section{Background and Motivation}

\subsection{Background}
3GPP has specified IoT services over non-terrestrial networks (NTN), extending the cellular IoT technique such as NB-IoT to operate over satellite links \cite{amatetti2022nb,liberg2021narrowband}. Unlike the terrestrial case, IoT-NTN operates under fundamentally different physical and network conditions, including long propagation delays and stringent power constraints. Among the various NTN architectures, geostationary earth orbit (GEO) satellites represent a particularly attractive option for IoT connectivity due to their wide coverage and fixed ground footprint, but also introduce unique challenges for protocol design and system operation.

\begin{figure}[t]
  	\centering
  	\includegraphics[width=8.5cm]{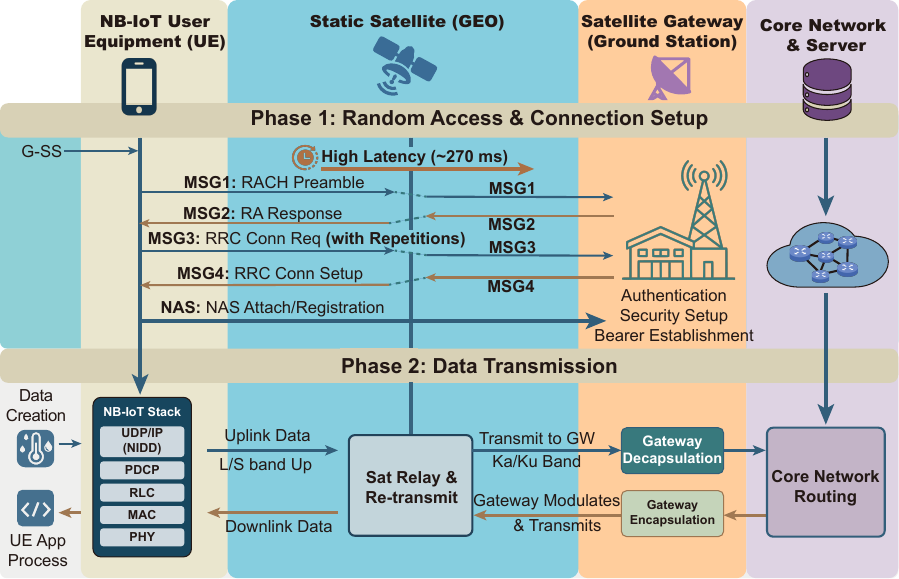}
	\caption{The uplink establishment process and measurement framework in NB-IoT NTN.}
    \label{process1}
\end{figure}

In a GEO IoT-NTN system, IoT devices communicate with a satellite located at approximately 35,786 km above the Earth, which relays traffic to a gateway connected with the terrestrial core network, as shown in Fig.~\ref{process1}. From the perspective of the 3GPP core, the satellite access network largely emulates a conventional cellular radio access network, enabling reuse of existing NB-IoT and LTE-M procedures with limited modifications. However, the satellite link introduces a one-way propagation delay of roughly 120–140 ms, resulting in a round-trip time exceeding 500 ms when accounting for processing and backhaul delays. This considerable delay affects time-sensitive procedures such as random access, scheduling, and retransmissions.


As described in \cite{yang2020understanding}, the conventional radio access procedure of an uplink transmission mainly consists of six steps:(1) The ground terminal initiates a contention-based Random Access (RA) by sending a Random Access Preamble (MSG1) to the satellite; (2)The satellite replies with a Random Access Response (MSG2), which allocates resource for the terminal to send subsequent requests; (3) The ground terminal sends a Scheduled Transmission (MSG3) to request resource needed for data transmission while running a Contention Resolution Timer (CRT); (4) The satellite resolves contention, allocates resource, and then notifies the ground terminal by sending a Contention Resolution (MSG4). If the ground terminal receives MSG4 before CRT timeout, the RA procedure completes successfully and the ground terminal sets up a Radio Resource Control (RRC) connection to the satellite. Otherwise, the terminal re-initiates RA; (5) After data transmission, the ground terminal enters a passive reception mode, listening to possible downlink packets until the RRC connection is released; (6) Finally, the core network sends the RRC connection release to the ground terminal to mark the end of a packet transmission cycle.

Rather than designing a new satellite-specific IoT stack, 3GPP reuses existing cellular mechanisms, and introduces only incremental enhancements to support NTN operation. These include extended timing advance ranges, modified random access procedures, relaxed latency constraints, and optional support for transparent or regenerative payloads. While this mechanism accelerates standardization and deployment, it also raises important questions about how well terrestrial-designed protocols perform under extreme satellite conditions—particularly for GEO systems where delay is orders of magnitude larger than that in terrestrial networks.

\subsection{Motivation}
\label{sec: Motivation}
In recent years, 3GPP Rel-17/18 have defined adaptations for operating NB-IoT over Non-Terrestrial Networks (NTN), accelerating the deployment of IoT-NTN and opening the possibility of global IoT connectivity via satellite systems. Among different NTN architectures, geostationary Earth orbit (GEO) satellites play a critical role due to their wide coverage, stable beams, and established infrastructure. However, despite standardization progress and growing industrial interest, the real-world behavior of 3GPP IoT-NTN over GEO remains largely unexplored. 


Measuring real-world performance of IoT-NTN over GEO satellites is essential and thus lays a good foundation for the future deployment and implementation of NB-IoT NTN. Actually, there exists available measurement work on IoT-NTN through simulations \cite{mea,mea1}. However, GEO satellite systems introduce uniquely demanding link characteristics that remain insufficiently understood through simulations or analytical models alone, as illustrated below.

\begin{itemize}
    \item The fundamental reason stems from the extreme propagation delay of GEO links, where the round-trip time typically exceeds 500 ms. Such delay is orders of magnitude larger than those in terrestrial cellular networks and directly affect core protocol procedures, including random access, RRC state transitions, HARQ operation, and higher-layer transport protocols. While 3GPP specifications define extended timers and NTN-specific adaptations, these parameters are derived from analytical assumptions and controlled laboratory emulation, leaving their effectiveness under real satellite deployments uncertain.
    
    \item Moreover, NB-IoT was not originally designed for GEO-scale propagation. Its synchronization mechanisms and control-plane procedures were optimized for terrestrial environments with negligible latency and tight timing constraints. In GEO IoT-NTN, low-cost IoT devices must cope with long timing advance, oscillator drift, asymmetric uplink/downlink paths, and satellite gateway queuing—effects that are difficult to model accurately and often absent from simulation studies. As a result, the actual reliability, latency, and energy efficiency experienced by IoT devices in operational GEO networks remain unknown. 


\end{itemize}

These factors motivate the need for a measurement-driven study of 3GPP IoT-NTN over GEO. From a system and networking perspective, GEO based IoT-NTN offers a compelling yet under-explored operating point. Understanding how standardized IoT protocols behave in such an environment requires empirical measurement and deployment experience, especially with respect to delay, energy consumption, and reliability. By conducting real-world experiments on operational or prototype GEO satellite systems, we can expose protocol behaviors that are invisible in simulation, validate (or challenge) standard assumptions, and provide actionable insights for future NTN design and deployment.



\section{End-to-end Delay}
We first measure the End-To-End (E2E) delay under different conditions such as different payload size, traffic interval, and concurrency level, to verify the adaptation capability of NB-IoT protocol stack in GEO NTN scenarios. Featured with ultra-long Round-Trip Time (RTT) and remote core network, the E2E delay is expected to be considerably prolonged compared to the terrestrial case. 

\textbf{Overview.} The E2E delay is measured by running the ping command, which starts from the ground NB-IoT terminal, passing through the GEO satellite, ground station, core network, and finally arriving at the cloud server located at Hong Kong. The ping command is executed 30 times for each configuration. Here, we measure this E2E delay based on two different NB-IoT NTN service providers--Skylo in America and Tiantong in China. In Skylo, the GEO satellite refers to INMARSAT 4-F1 located at $(4.41^\circ, 178.17^\circ)$ with the altitude of 35794 km, while its ground station locates at Hongkong. However, its core network is deployed in the United States. Correspondingly, Tiantong GEO satellite locates at $(0.25^\circ, 101.44^\circ)$ with the altitude of 35772 km. Meanwhile, its ground station and core network are at Xi'an and Beijing, respectively. 

\subsection{Under different payload size}
\label{E2E}

\begin{figure}
    \centering
    \includegraphics[width=0.8\linewidth]{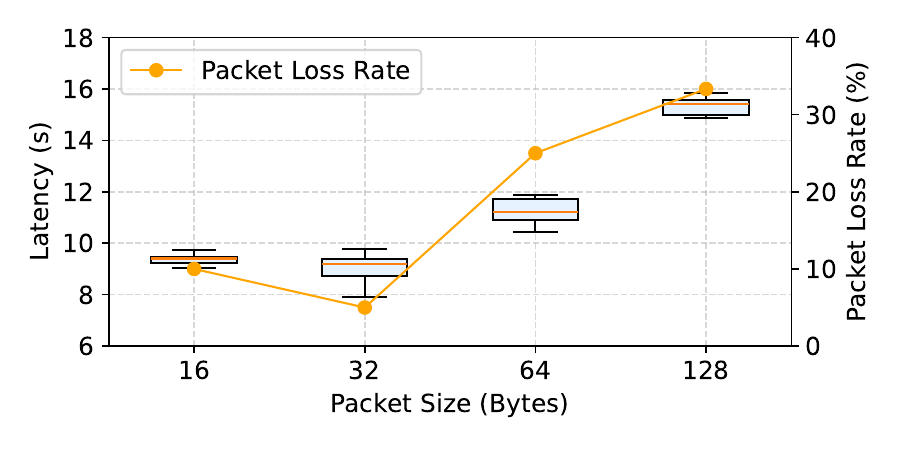}
    \caption{E2E latency and packet loss rate under different payload sizes in Skylo-based NB-IoT NTN. A clear latency inflection point appears when the payload exceeds 32 bytes, after which latency increases approximately linearly due to uplink transport block fragmentation and serialized NPUSCH transmissions.}
    \label{ping1}
\end{figure}

\textbf{Skylo.} As shown in Fig.~\ref{ping1}, in Skylo NB-IoT NTN, the E2E delay remains nearly constant for small payloads yet increases sharply once the payload size exceeds a threshold, exhibiting an approximately linear increase trend. Specifically, when the payload size is less than 32 bytes, E2E delay remains relatively stable and centered around 9 seconds. However, when the payload size exceeds 32 bytes, a clear delay inflection point emerges. For example, the median delay increases to approximately 11–12 seconds at 64 bytes, and further rises to over 15 seconds at 128 bytes. Notably, the delay increase beyond this threshold is approximately linear with respect to payload size. 

In addition to delay inflation, it also shows a monotonic increase of packet loss as the payload size grows. This indicates that larger payloads not only require more uplink fragments but are also more vulnerable to transmission failures, further exacerbating end-to-end performance degradation in NB-IoT NTN.

\begin{figure}
    \centering
    \includegraphics[width=0.8\linewidth]{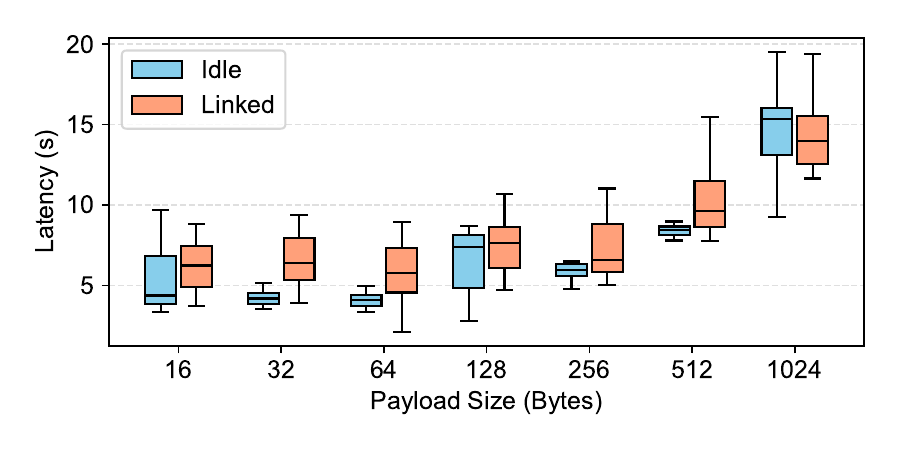}
    \caption{End-to-end delay under different payload size at both RRC idle and linked states in Tiantong-based NB-IoT NTN.}
    \label{ping2}
    \vspace{-\baselineskip}
\end{figure}

\textbf{Tiantong.} Figure~\ref{ping2} presents the E2E delay performance under different payload size in Tiantong NB-IoT NTN, while simultaneously distinguishing between the RRC idle and linked states. For payload size ranging from 16 to 64 bytes, the E2E delay remains relatively low and stable at both states, with the idle state exhibiting slightly lower median latency. As the payload size increases beyond 128 bytes, E2E delay rises steadily at both states, with a noticeably steeper increase at the linked state. At payload sizes of 512 and 1024 bytes, the median E2E delay exceeds 10 seconds and continues to grow rapidly coupled with increased variance. 

Interestingly, we also find that the delay at the idle state is shorter than the linked state, which is in contrast to the common sense that the delay at the linked state should be shorter than the idle case since the linked state does not require the access procedure. The reason is that when sending small data packets at the idle state, it usually does not require the establishment of a complete dedicated data channel (DRB). It may utilize MSG3 to directly carry and send data, thereby achieving fast transmission of small packets. However, this mechanism is unreliable and prone to conflicts incurred by competition for resources, leading to significant delay fluctuations (i.e., high standard deviation). In the linked state, a complete data transmission link needs to be established. For small packets, the high signaling overhead results in an E2E delay that is larger than that in the idle state. However, the advantage is that the transmission is very stable. When transmitting large packets like 1024 bytes, the linked state begins to catch up with or even slightly surpasses the idle state.

\textbf{Observation$\&$conclusion.} Across the above two commercial NB-IoT NTN systems (i.e., Skylo and Tiantong), the End-To-End (E2E) delay remains stable for small payloads but increases sharply once the payload size exceeds a system-specific threshold, exhibiting a near-linear increase trend that is independent of operator and RRC state. This behavior can be attributed to the limited Transport Block Size (TBS) specified in NB-IoT NTN uplink transmissions. That is to say, due to conservative modulation and coding schemes, high repetition factors, and large timing guard intervals required in satellite links, a single NPUSCH transmission can only carry a small effective payload. Once the payload exceeds this limit, it must be fragmented into multiple sequential NPUSCH transmissions, which is in accordance with the regulations in 3GPP TS 36.523-2 \cite{johansson20133gpp}. When the payload size is relatively small within a single uplink NPUSCH transmission, the latency is dominated by fixed control-plane procedures and satellite propagation delay. Otherwise, each additional fragment incurs extra scheduling delay, repetition overhead, and RTT. Since these uplink transmissions are serialized rather than pipelined, the accumulated delay increases proportionally with the number of required transport blocks, leading to the linear latency increase as observed.

Simultaneously, there exist two main differences between Skylo and Tiantong. First, according to these measurement results, the TBS threshold is around 32 and 128 bytes in Skylo and Tiantong, respectively. Specifically, the TBS threshold set in Skylo is much smaller than the specifications (i.e., 1000 bits) in 3GPP TS 36.523-2, while Tiantong is consistent with the standard regulation. Given that these commercial satellite systems are closed-source, we can not obtain the detailed settings in physical layer. The most possible reason is that Tiantong NB-IoT NTN remains at the under-constructed stage, while Skylo NB-IoT NTN has already been applied for IoT applications, which indicates 32 bytes is a more suitable TBS threshold in actual deployments. 

Secondly, we observe that the end-to-end delay based on Skylo is higher than that based on Tiantong even the packet size in both cases are smaller than 32 bytes. To investigate, we make a delay breakdown. Note that we can not utilize the measurement tool like traceroute on ground IoT terminals for delay breakdown, thus unable to obtain the detailed delay for each segment within the whole path. At a high level, the packet is first uploaded to the GEO satellite, and then passing through the ground station, core network, and finally arriving at the server located at Hong Kong. However, in Skylo and Tiantong NB-IoT NTN over GEO, the transmission delay between the ground terminal and GEO satellite can be approximated as 600 ms. In Skylo, the transmission delay between the ground station and core network (located at America) is high up to 300 ms, while this delay is only 40 ms in Tiantong. It exactly explains why the end-to-end delay based on Skylo is higher than that based on Tiantong.

\begin{figure}
    \centering
    \includegraphics[width=1\linewidth]{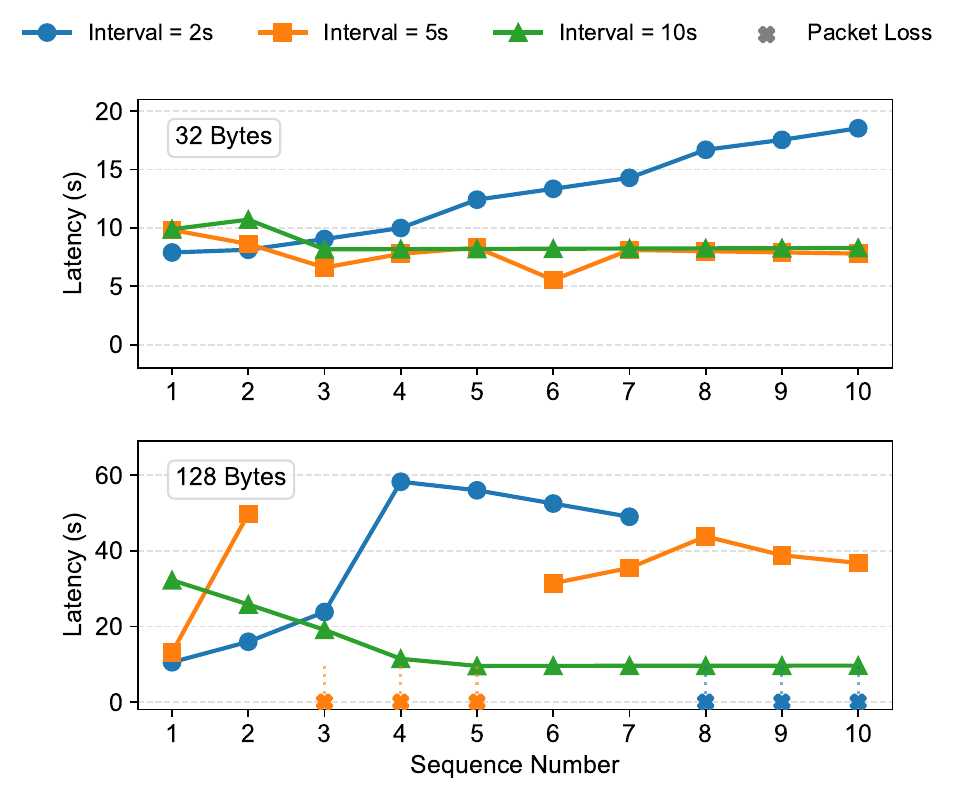}
    \caption{End-to-end delay under different traffic interval.}
    \label{new_uplink2}
\end{figure}

\subsection{Under Different Traffic Interval}
Next, to investigate how reporting frequency shapes NB-IoT NTN behavior, packets are conveyed serially with different intervals.

Figure~\ref{new_uplink2} describes the E2E delay with samples sorted by latency under different intervals, where a clear interval-dependent trend emerges. With a short interval of 2 seconds, E2E delay increases and eventually exceeds 17 seconds. It indicates that even for packets that are ultimately received successfully, aggressive transmission pacing leads to increasing queuing and retransmission delays as contention accumulates. In contrast, longer intervals such as 10 seconds exhibit much flatter curves, where E2E delay remains nearly constant and suggesting a more stable access process. When the payload size increases to 128 bytes, we can observe some packet loss when setting a short interval, which implies that adopting a smaller payload size is more suitable for NB-IoT NTN over GEO and thus consistent with the analysis in subsection~\ref{E2E}.




\begin{figure}
    \centering
    \includegraphics[width=0.8\linewidth]{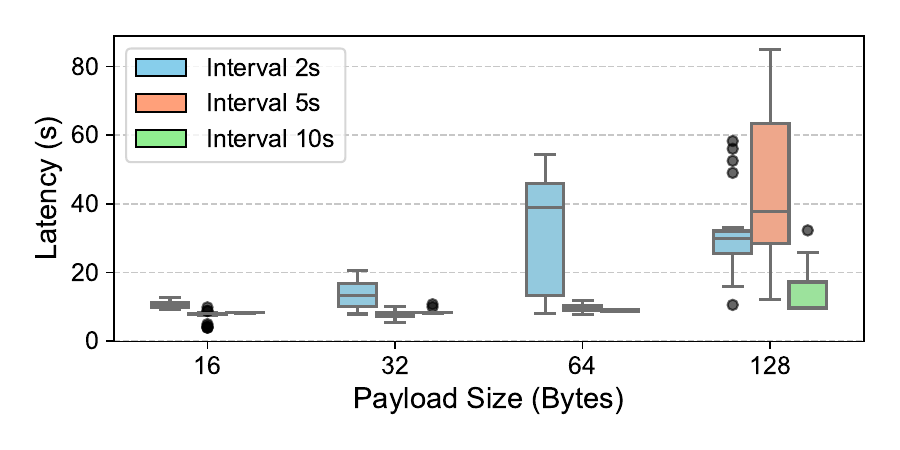}
    \caption{End-to-end delay under different traffic interval and payload size.}
    \label{latency1}
    \vspace{-\baselineskip}
\end{figure}


To further investigate the integrated impact of payload size and traffic interval, figure~\ref{latency1} illustrates the E2E delay observed under different payload sizes (16–128 bytes) and transmission intervals (2 s, 5 s, and 10 s). Overall, E2E delay increases with payload size. However, for small payloads (16–32 bytes), E2E delay remains relatively stable across different intervals. This suggests that under lightweight traffic conditions, E2E delay is primarily dominated by the access procedure (e.g., random access and control signaling) rather than the actual payload size. When the payload increases to 64 and 128 Bytes, E2E delay increases rapidly and exhibits greater variability, particularly under a short interval of 2 seconds, where both the median latency and dispersion increase substantially. It indicates that under heavy traffic conditions, frequent transmission intensifies access contention and scheduling delay, leading to degraded performance. In contrast, a long interval of 10 seconds effectively mitigates network load and results in lower and more stable E2E delay. Furthermore, the presence of outliers under large payloads and short intervals highlights the impact of repeated transmissions and inherent long delay of GEO links, which together introduce considerable performance variability in NB-IoT NTN systems.

\textbf{Observation$\&$conclusion.}
The divergence between short and long intervals highlights that E2E delay is mainly affected by MAC and scheduling dynamics rather than propagation delay. While all curves share a common lower bound set by GEO RTT and core-network processing, only the short-interval cases experience rapid delay inflation due to access congestion and HARQ/NPUSCH retries. \textbf{\textit{These results imply that ensuring reliable and predictable latency for successful packet delivery requires conservative transmission intervals, reinforcing that pacing control is a key lever for stabilizing performance.}}

\subsection{Under Different Concurrency Level}
 
\textbf{Overview.} Finally, we have measured the E2E delay under concurrent transmission scenarios, which frequently occur in IoT-NTN. To mimic the concurrent transmission scenario, we have deployed ten NB-IoT nodes (i.e., JMD47) in an outdoor area, and control them to convey the sensed information to the GEO satellite simultaneously. The concurrency level ranging from 1 to 10 denotes the number of concurrently transmitted devices. Meanwhile, this kind of measurement is based on the Skylo network.    

\begin{figure}
    \centering
    \includegraphics[width=0.8\linewidth]{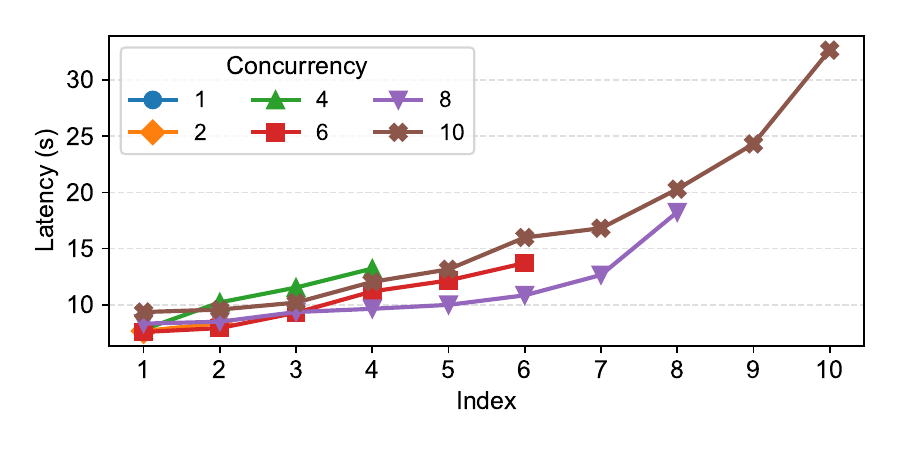}
    \caption{The E2E delay of terminals under different concurrency levels.}
    \label{concurrent1}
\end{figure}

\subsubsection{Concurrency=1} As shown in Fig.~\ref{concurrent1}, the E2E latency in NB-IoT NTN over GEO maintains a high level up to several seconds even when there exists only one single ground device attempting to connect with the GEO satellite, which is much higher compared to the terrestrial case. We can obtain the following observation: \textbf{\textit{Even under single-device and low-traffic conditions, NB-IoT over GEO exhibits a non-negligible baseline latency of multiple seconds, primarily due to long satellite round-trip propagation and non-co-located core network processing.}} 

\subsubsection{Concurrency$\textgreater$1} NB-IoT NTN over GEO is fundamentally bottlenecked by shared control-plane and satellite resources, parallel communication exposes non-linear congestion, synchronization, and energy collapse effects that cannot be observed in single-terminal experiments. To investigate the concurrent transmission scenarios, we have deployed ten ground terminals in our campus. 

Figure~\ref{concurrent1} illustrates the E2E delay of NB-IoT devices under different concurrency level. As the device number increases from 1 to 10, the entire delay curve shifts upward and becomes increasingly skewed, indicating both higher average latency and a pronounced tail. Devices with lower indices experience only moderate latency increase, while high-index devices suffer a sharp increase with tail exceeding 30 seconds under high concurrency. This behavior stems from the ultra-long RTT combined with NB-IoT’s contention-based random access and sequential scheduling, where RACH collisions, delayed RAR/uplink grants, and retransmissions accumulate as more devices attempt to access the network simultaneously. \textbf{\textit{In NB-IoT over GEO satellites, E2E delay increases super-linearly with the number of concurrent UEs, even when each UE transmits small payloads at low data rates.}}


\begin{figure}
    \centering
    \includegraphics[width=0.8\linewidth]{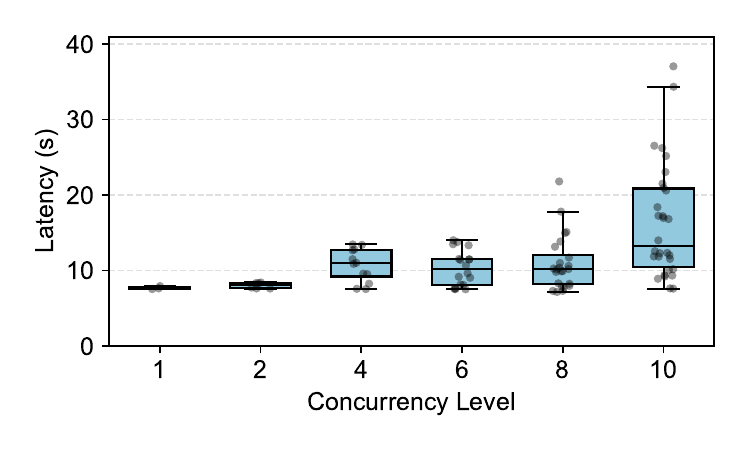}
    \caption{The E2E delay distribution under different concurrency level.}
    \label{concurrent2}
    \vspace{-\baselineskip}
\end{figure}

Figure~\ref{concurrent2} shows the E2E delay distribution when increasing the concurrency level. When concurrency is low (1–2 devices), E2E delay is tightly concentrated around 8–9 seconds with small variance, which indicates a stable and predictable access process dominated by the fixed propagation delay. At the concurrency level beyond 6, the delay distribution becomes highly dispersed, with medians rising to roughly 10–15 seconds and upper whiskers extending beyond 35 seconds, revealing severe variability across devices. This widening distribution highlights the strong tail-latency effect inherent within NB-IoT NTN under intensive contention. While a subset of devices can still complete access within tens of seconds, others experience significantly prolonged delays due to RACH collision, delayed scheduling, and retransmissions amplified by the long RTT. This widespread distribution indicates queue buildup and scheduling delays in the NB-IoT access and gateway/core-network pipeline, rather than additional propagation delay.

\textbf{Observation$\&$conclusion.} \textbf{\textit{This behavior highlights that scalability is primarily limited by access and network-side contention, making concurrency-induced tail latency a critical performance bottleneck for delay-sensitive IoT applications.}}From an application standpoint, this implies that average or median latency alone is insufficient to characterize system performance. Instead, reliability-sensitive or synchronized applications are constrained by worst-case access times, underscoring the need for concurrency control, access staggering, or priority-based mechanisms to suppress tail latency in practical NB-IoT NTN deployments.

\section{Energy Consumption vs Delay}
In addition to the delay, another key feature of NB-IoT NTN is the promise of long battery life, up to 10 years. However, realizing this aim in NTN scenarios is challenging when considering the ultra-long transmission distance from the ground terminal to GEO satellites. Investigating how the power is spent and identifying the energy bottleneck is of great importance to the design and large-scale implementation of NB-IoT NTN.  

We first investigate the relationship between the above E2E delay and energy consumption by building the uplink latency model, as shown below.    
\begin{equation}
    T_{E2E}=T_{sync}+T_{RA}+T_{RRC}+T_{UL}+T_{core},
\end{equation}

\noindent where $T_{sync}$ denotes the synchronization time and $T_{RA}$ represents the time for random access waiting and retrying. $T_{RRC}$ indicates the Round-Trip Time (RTT) for RRC connection. Finally, $T_{UL}$ and $T_{core}$ correspond to the packet delivery time from the ground terminal to GEO satellite and from GEO satellite to the core network, respectively. In particular, $T_{sync}$ equals to the multiplication number of synchronization attempts--$N_{sync}$ and synchronization window duration--$T_{sync,w}$. Meanwhile, $T_{RA}$ is determined by 
\begin{equation}
    T_{RA}=N_{RA}\times (T_{NPRACH}+T_{RTT}+T_{RAR}), 
\end{equation}

Note that in NTN scenarios, $T_{RTT}$ is much larger than $T_{NPRACH}$. $T_{RRC}$ actually equals to multi-round RTT superposition for conveying controlling signal, which is determined by the number of control signaling round trips--$N_{c}$ and $T_{RTT}$. According to the detailed analysis in Section~\ref{E2E}, $T_{UL}$ and $T_{core}$ can be neglected. Therefore, in GEO NTN scenarios, $T_{E2E}$ can be approximated as 
\begin{equation}
   T_{E2E}=N_{RA}\times T_{RTT}+N_{c}\times T_{RTT}, 
\end{equation}

Similarly, the energy consumption $E_{t}$ can be represented as 
\begin{equation}
    E_{t}=E_{RA}+E_{RRC}, 
\end{equation} 

\noindent where energy is primarily consumed by the random access and RRC connection process. Note that $E_{RA}$ can be further represented as the following equation.
\begin{equation}
   E_{RA}=N_{RA}(P_{TX}T_{NPRACH}+P_{RX}T_{wait}), 
\end{equation} 

In the NTN scenarios, $T_{wait}$ equals to $T_{RTT}$. Meanwhile, $E_{RRC}$ can be denoted as 
\begin{equation}
  E_{RRC}=P_{RX}T_{RRC}, 
\end{equation} 

Combining the above three equations, $E_{t}$ can be approximately represented as
\begin{equation}
\label{e1}
  E_{t}=P_{RX}T_{E2E}+N_{RA}\times P_{TX}T_{NPRACH}. 
\end{equation} 

Equation~\ref{e1} indicates that the dominant energy term scales linearly with end-to-end delay, reducing access repetitions and connection time simultaneously minimizes both latency and energy consumption, revealing a unique and favorable energy–latency tradeoff.

\subsection{Power consumption using different transmission schemes}
\begin{figure}
    \centering
    \includegraphics[width=1\linewidth]{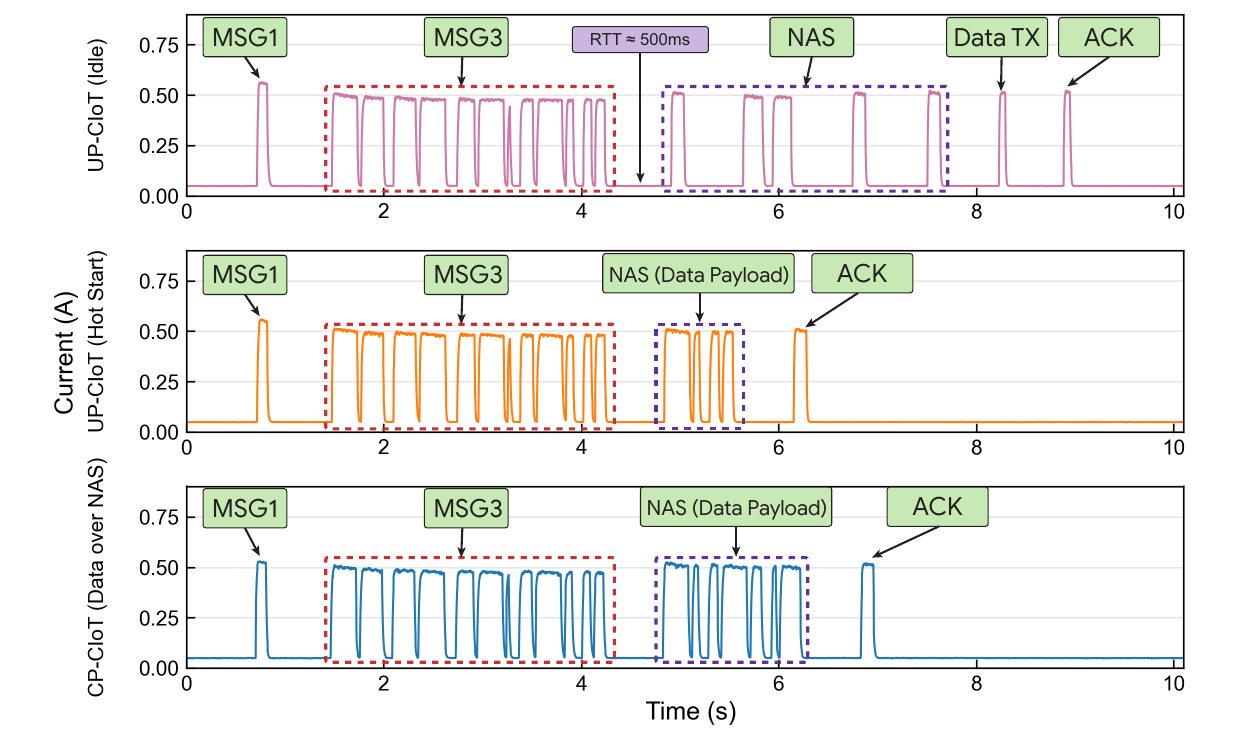}
    \caption{Power consumption trace during the connection process with different transmission schemes.}
    \label{energy1}
\end{figure}

Next, we explore how the energy is consumed over time, and the power trace is measured by the UT70 USB measurement device. As shown in Fig.~\ref{energy1}, the traditional scheme--UP-CIoT (Idle) requires a complete random access and RRC connection establishment process, which involves six handshake interactions. Each interaction necessitates listening and waiting for a response from the GEO satellite, and thus bringing a propagation delay of 500 ms. This cumulative effect of "interaction-waiting" extends the overall end-to-end delay ranging from 8 to 9 seconds. During these periods, the radio frequency module cannot enter the deep sleep mode (PSM), and even in a listening state, its accumulated tail current represents a non-negligible energy consumption. Consequently, the long RTT of around 500ms, which includes path delay and processing delay, undermines the performance of conventional cellular handshake protocols.

\begin{figure}
    \centering
    \includegraphics[width=1\linewidth]{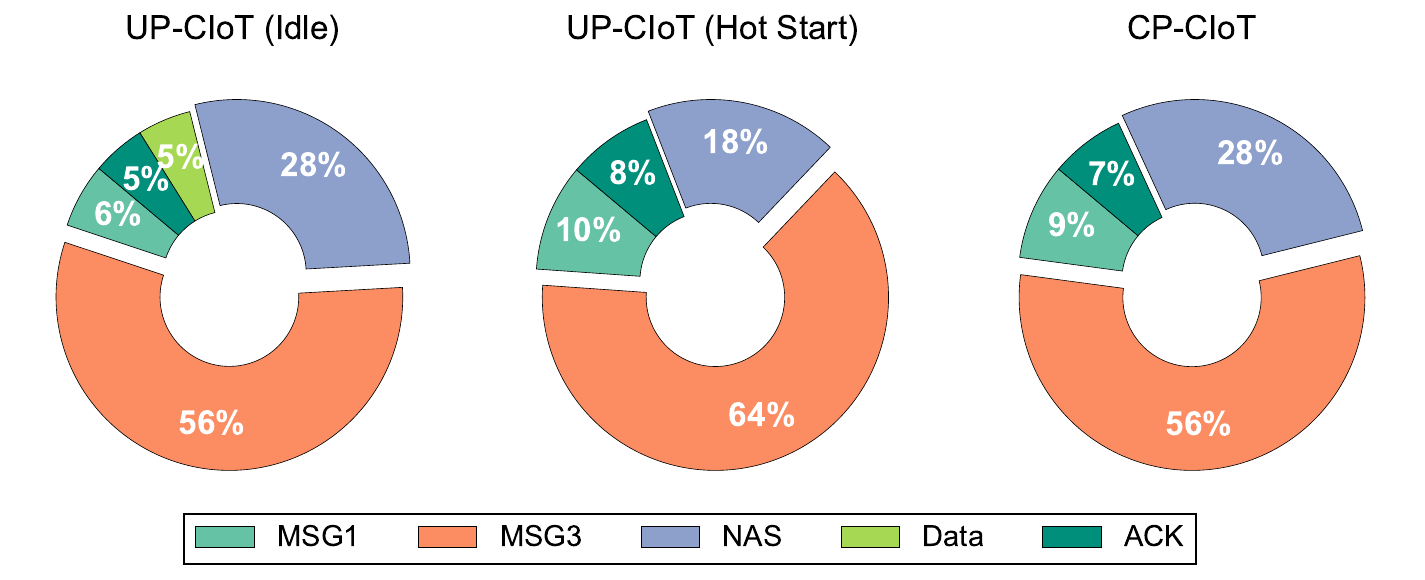}
    \caption{The energy consumption during different steps.}
    \label{energy2}
\end{figure}

Correspondingly, as described in Fig.~\ref{energy2}, the data transmission phase, which is actually used for data transmission, consumes only 5\% of the total energy for uplink packet transmission. However, the energy produced by the repeated transmission of MSG3 for random access accounts for 56\% of the total energy, which is more than half. It demonstrates that blind retransmission to overcome the huge free-space loss in GEO satellite communication leads to significant energy waste. At the same time, the energy consumption in the NAS signaling interaction phase accounts for 28\%, closely following the energy consumption in MSG3. This confirms that in the NB-IoT NTN over GEO, the traditional network attachment, authentication, and bearer establishment mechanisms appear overly cumbersome. Most of the system's energy is used to maintain the operation of the underlying protocols, rather than for effective data transmission.

To overcome the aforementioned issues of signaling redundancy and long RTT, 3GPP introduced the CIoT optimization design--CP-IoT(Data over NAS) with its core concept being "Piggybacking". Its key is to directly involve encapsulating small data packets within NAS signaling for transmission, thereby completely eliminating the need to establish a separate user plane data bearer (DRB). Fig.~\ref{energy1} describes the optimized current waveform for packet delivery over the control plane. Compared with UP-CIoT(idle), it can be observed that the number of NAS handshakes is reduced, and the overall delay is reduced from around 8 to 6 seconds and thereby reducing the delay. However, in terms of energy consumption, signaling overhead still accounts for the vast majority of energy consumption, as shown in Fig.~\ref{energy2}. This is because CP-CIoT is optimized for RRC connections and does not optimize the repeated transmission of MSG3 during the random access process. 

Compared to another transmission scheme--UP-CIoT(hot start), the optimization effect of CP-IoT(Data over NAS) does not surpass this method because the UP-CIoT (hot start) utilizes the RRC suspend/resume mechanism for allowing the terminal to retain the network context after the previous transmission, thereby skipping some cumbersome authentication and bearer establishment processes. Furthermore, the UP-CIoT(hot start), which preserves the RRC connection, even slightly outperforms CP-CIoT(Data over NAS) in terms of latency and energy consumption.

Therefore, in actual NB-IoT NTN over GEO, if the transmission frequency is extremely low such as once a day and the network context cannot be retained for a long time, the conventional CP-CIoT can be adopted. If the transmission is intermittent or has a short cycle (which can be maintained within the window where signaling does not timeout), UP-CIoT (Hot Start) is the optimal design.

\textbf{Observation$\&$conclusion.}According to the measured power consumption traces, the NB-IoT link connection procedure is characterized by multiple high-power transmission bursts of MSG3 and NAS. The NPRACH-based random access phase dominates the total energy consumption and is determined by the number of repeated transmission and round-trip delay, accounting for nearly half of the overall connection energy (around 56\%). Due to the extended round-trip delay and repeated uplink transmissions in NTN, the connection establishment cost is significantly higher than that of terrestrial NB-IoT \cite{yang2020understanding}, while the actual data transmission contributes only a minor portion. Therefore, the low-power design of traditional terrestrial NB-IoT fails in the NTN case. \textbf{\textit{Compared to terrestrial NB-IoT, the energy overhead per connection is increased by approximately 3–5 times, severely limiting the battery life of low-power IoT devices in NTN scenarios.}}

\begin{figure}
    \centering
    \includegraphics[width=1\linewidth]{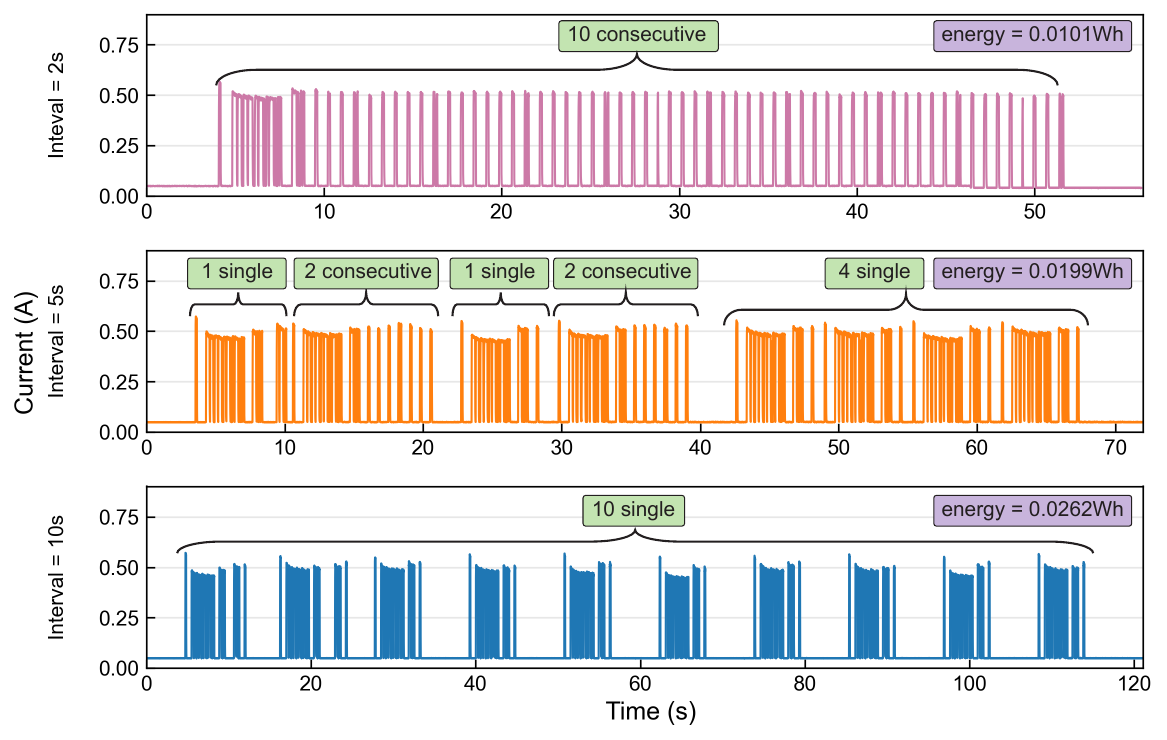}
    \caption{The power trace with different traffic intervals when sending 10 consecutive packets.}
    \label{interval1}
\end{figure}

\subsection{Power consumption trace with different traffic interval}
Next, we investigate the power consumption when transmitting multiple packets. As shown in Fig.~\ref{interval1}, the energy consumption for sending 10 packets is 0.0101 Wh, 0.0199 Wh, and 0.0262 Wh, respectively, corresponding to the three intervals (2, 5, 10 s). The total energy consumption for a 10-second interval is almost twice compared to that of a 2-second interval. This is because the ground terminal remains in RRC connected mode due to frequent transmissions in the case of 2-second interval, and subsequent data packets can be directly sent through PUSCH, thus eliminating the cumbersome handshake process. However, in the 10-second interval, the terminal enters a sleep or RRC idle state after sending the packet, and the next transmission must be "woken up" and go through a complete random access signaling interaction. This part of the "signaling overhead" used to establish a connection consumes the majority of the battery power. 

However, in the waveform with a 5-second interval, the device exhibits an extremely unstable state: sometimes it can maintain a connection and sometimes it needs to randomly re-access, which is in contrast to the absolute continuity of 2 seconds and the absolute disconnection of 10 seconds. This phenomenon reveals that the RRC inactivity timer threshold is most likely around 5 seconds. Due to the long propagation delay in NB-IoT NTN over GEO, coupled with network jitter and protocol stack processing delay, the arrival of the next data packet happens to be around the timer timeout point. To illustrate, when the network is stable, the packet arrives before the timer timeout and the ground station considers the device to be still active, thus maintaining the connection; if there exists a slight delay and the timer just times out to release the connection before the packet's arrival, forcing the device to re-initiate MSG1-4. 

\textbf{Observation$\&$conclusion.}Therefore, \textbf{\textit{directly copying the timer parameters from the terrestrial network to the satellite IoT scenario leads to uncontrollable state transitions.}} In practical deployment for IoT-NTN, it is necessary to re-calibrate and extend the inactivity timer on the network side according to the RTT characteristics of GEO satellites, or avoid this kind of traffic interval that easily triggers jitter at the application layer.


\subsection{PSM}
\begin{figure}
    \centering
    \includegraphics[width=0.8\linewidth]{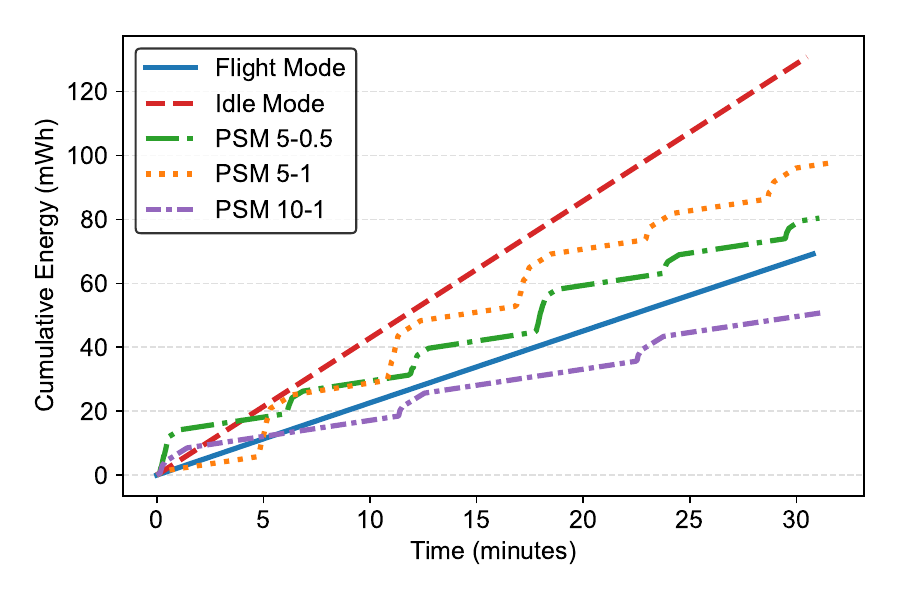}
    \caption{Cumulative energy over time in PSM.}
    \label{psm1}
\end{figure}

Here, we note that ground NB-IoT terminals will eventually enter the Power Saving Mode (PSM) after completing packet delivery. From the time-accumulated distribution shown in Fig.~\ref{psm1}, the energy consumption in PSM of the IoT-NTN scenario exhibits a stepped pattern, reflecting the non-negligible impact of GEO satellite delay on energy consumption. Taking PSM 10-1 as an example, the gentle plateau period in the curve corresponds to the stable and low noise floor during PSM's deep sleep, while the periodic steep jumps map the active processes of terminal wake-up, downlink synchronization, random access, and TAU signaling interaction. Here, PSM 10-1 denotes two timers in PSM, where 10 represents T3324 timer--how long the UE stays reachable in idle mode before entering PSM and 1 indicates T3412 timer--the maximum duration of the PSM cycle. Due to the ultra-long propagation delay of GEO satellites, the radio front-end and baseband processing unit are forced to suspend for a long time and maintain a high-power state during the completion of MSG1-4 handshake and waiting for network side confirmation. This physical waiting time caused by the ultra-long RTT directly translates into a heavy energy penalty, resulting in a wide energy consumption step that is different from the terrestrial NB-IoT network. Under the NB-IoT NTN over GEO architecture, physical layer delay will inevitably lead to a high energy consumption.



This stepped consumption pattern, observed from a system-level perspective, reveals that PSM fundamentally reshapes NB-IoT NTN traffic into a bursty and access-driven communication pattern. While PSM is highly effective in reducing idle energy consumption, its interaction with long round-trip time amplifies access latency and variability. In NB-IoT NTN deployments, PSM configuration must be carefully aligned with application traffic patterns; otherwise, frequent wake-ups can negate latency predictability and degrade overall efficiency. The results suggest that PSM-aware traffic shaping and access optimization are essential to achieving a practical balance between energy efficiency and latency in NB-IoT NTN systems.

\section{Implications for NB-IoT NTN Design}
To conclude, the impact on E2E delay and energy consumption can be divided into two segments--before and after access. Before access, we should minimize the access competition. After access, we can minimize retransmission of MSG3 (i.e., retransmission number for resource request).

\begin{figure*}[htbp]
    \centering
    \begin{minipage}[t]{0.3\textwidth}
        \centering
        \includegraphics[width=1.05\linewidth,height=5cm,keepaspectratio]{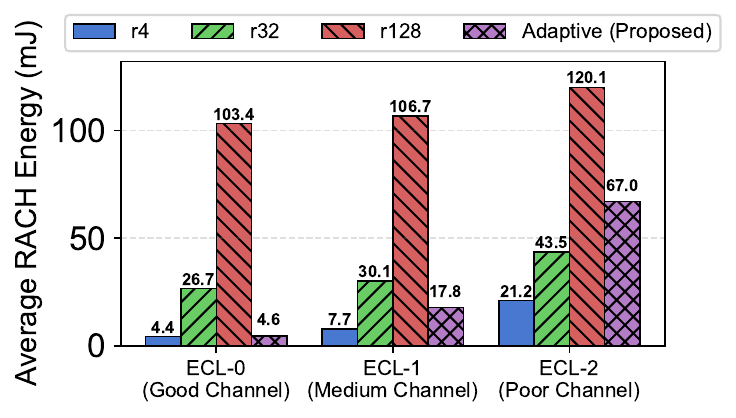}
        \caption{Energy of RACH based on different schemes operating at different ECL levels.}
        \label{adaptive1}
    \end{minipage}\hspace{0.02\textwidth}
    \begin{minipage}[t]{0.3\linewidth}
        \centering
        \includegraphics[width=1.05\linewidth,height=5cm,keepaspectratio]{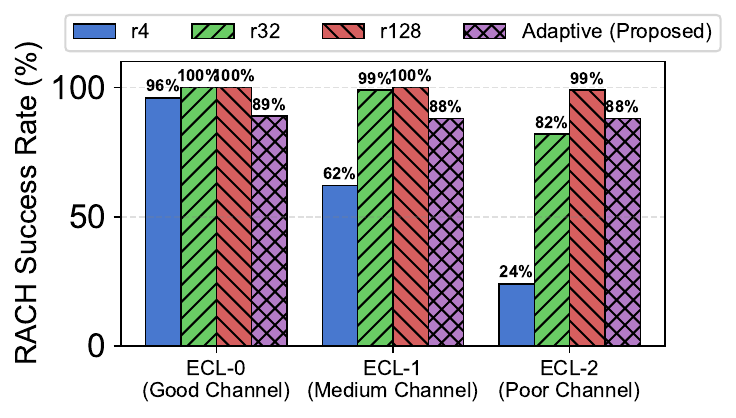}
        \caption{The RACH success rate under different ECL levels.}
        \label{adaptive2}
    \end{minipage}\hspace{0.02\textwidth}
    \begin{minipage}[t]{0.3\linewidth}
        \centering
        \includegraphics[width=1.05\linewidth,height=5cm,keepaspectratio]{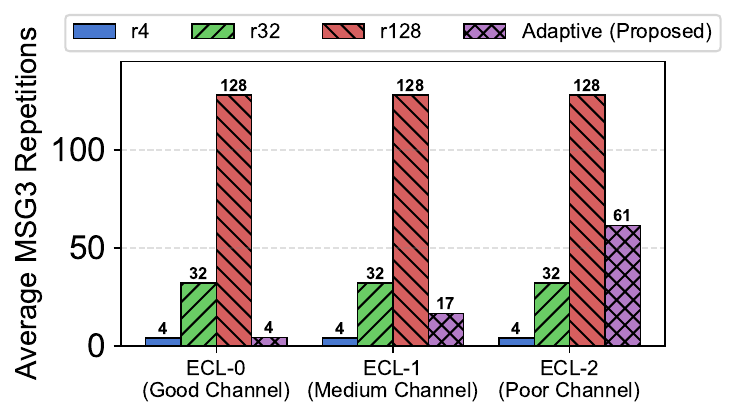}
        \caption{The assigned number of MSG3 retransmission under different ECL levels.}
        \label{adaptive3}
    \end{minipage}\hspace{0.02\textwidth}
    \begin{minipage}[t]{0.23\linewidth}
        \centering

    \end{minipage}
    \vspace{-\baselineskip}
\end{figure*}

Next, we propose corresponding optimal strategies to reduce the end-to-end delay and energy consumption. Specifically, to minimize the retransmission number of MSG3, we design an adaptive retransmission scheme for MSG3 to avoid unnecessary MSG3 delivery, the intuition behind which is to provide fine-grained control for MSG3 transmission based on the channel conditions. However, there only exist three coarse-grained selections for the retransmission number of MSG3 based on three ECL levels (e.g., ECL0, ECL1, and ECL2) in current NB-IoT NTN design \cite{zhang2024performance}. Here, ECL0, ECL1, and ECL2 denote the good, medium, and poor channel conditions, which respectively require the retransmission number of 4, 32, and 128. The designed adaptive scheme is to achieve real-time power estimation of MSG3 and accurately determine the required retransmission number by exploiting MSG1 delivery, since they share a coherent channel due to the little time gap between them, as shown in Fig.\ref{process1}. Actually, this scheme is lightweight and simple, which first estimates the reception power--$P_{rx,msg1}$ of MSG1 and then derives the power gain needed for receiving MSG3 successfully. Finally, the retransmission number of MSG3 can be calculated according to the estimated power gain, as described in the following Equation. 

\begin{equation}
    N_{rep}=ceil(10^{(P_{required}-P_{rx,msg1}+margin)/10}).
\end{equation}

\noindent where $P_{required}$ represents the required power gain for MSG3 reception and $margin$ denotes the power redundancy for reliable reception. $ceil$ indicates rounding up. We have conducted trace-driven simulations based on NS3 LENA \cite{schubert2022ns} to evaluate the performance of the proposed adaptive scheme. As shown in Fig.~\ref{adaptive1}, for one successful RACH, the adaptive scheme brings a similar energy consumption with r4 (i.e., the retransmission number is equal to 4) under good channel conditions, which is $6\times$ and $23\times$ less than r32 and r128, respectively. However, when operating under moderate channel conditions, the adaptive scheme consumes $44\%$ and $83\%$ less power than r32 and r128, respectively. Finally, under the worst channel conditions which requires the retransmission nymber of r128, the RACH process based on the adaptive scheme consumes $44\%$ energy less than r128 for MSG3 reception.   

As shown in Fig.~\ref{adaptive2}, when under good channel conditions, the RACH success rate based on r4 is $96\%$, while the success rate based on r32 and r128 can reach $100\%$. In this case, the success rate of the adaptive scheme is $89\%$, which exhibits a little less than r32 and r128. Yet, it consumes much less power ($6\times$ and $23\times$ reduction) compared to both r32 and r128. Thus, the adaptive scheme achieves a much higher efficiency in comparison to the available schemes when considering both E2E delay and energy consumption.

Figure~\ref{adaptive3} describes the retransmission number of MSG3 under different ECL levels. Compared to the available scheme for MSG3 retransmission (i.e., 4, 32, and 128 set under ECL1, ECL2, and ECL3, respectively), the adaptive scheme selects 4, 17, and 61 as the MSG3's retransmission number, which accurately match the channel requirements of each ECL level as the channel variation increases reasonably. Consequently, the proposed fine-grained adaptive scheme can considerably reduce both the delay and energy consumption in comparison with the available adjustment strategy in NB-IoT NTN.

Subsequently, to mitigate access competition, we propose two novel schemes--MIDA (Multiplicative Increase and Additive Decrease) and ACB-aware access mechanism. 

As highlighted above, PSM transforms latency from a channel-dominated metric into an access- and traffic-driven phenomenon in NB-IoT NTN over GEO. Applications should therefore generate traffic at intervals that are longer than or synchronized with the PSM cycle, allowing packets to be transmitted within a single active window and avoiding repeated access delays. Thus, we introduce MIDA, a lightweight T3324 controller for PSM timer inspired by the TCP congestion control logic. By tracking per-device delivery outcome, a lost packet doubles T3324 (fast recovery), while three consecutive successful deliveries shrink T3324 by a fixed step (slow probe). The asymmetric update rule mirrors TCP's multiplicative decrease and additive increase, but operates in the opposite direction—expanding the window under high delay rather than under congestion. We conduct simulations to verify MIDA's high efficiency, where the delay consists of propagation delay, access delay, and queue delay. In the simulations, traffic follows a two-state MMPP coupled with idle arrival rate $\lambda=1$ pkt/min and burst rate $\lambda=10$ pkt/min. Meanwhile, the duration of idle and burst states equal to 30 min and 10 min, respectively, where the steady-state burst occupies $25\%$. Finally, we model energy based on actual NB-IoT chip measurements: setting 59.8 mW at deep PSM sleep state, 259.4 mW at active time, and 1679 mW during uplink transmission. Simulations are conducted covering 6 hours with three independent seeds.

\begin{figure*}[htbp]
    \centering
    \begin{minipage}[t]{0.3\textwidth}
        \centering
        \includegraphics[width=1.05\linewidth,height=5cm,keepaspectratio]{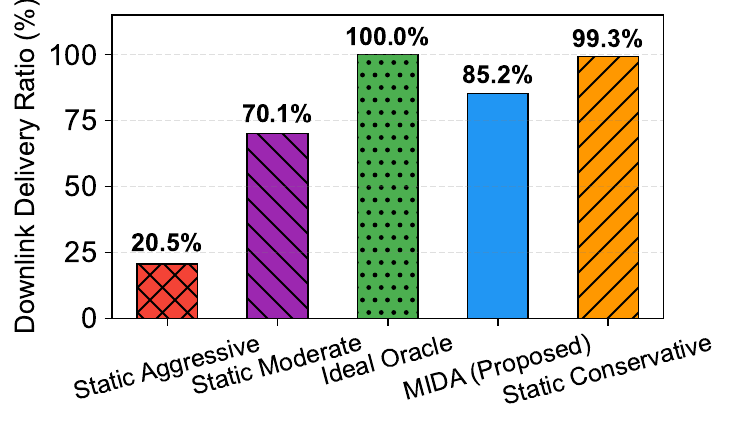}
        \caption{Downlink Delivery Ratio (DDR) based on different timer setting schemes in PSM.}
        \label{psm2}
    \end{minipage}\hspace{0.02\textwidth}
    \begin{minipage}[t]{0.3\linewidth}
        \centering
        \includegraphics[width=1.05\linewidth,height=5cm,keepaspectratio]{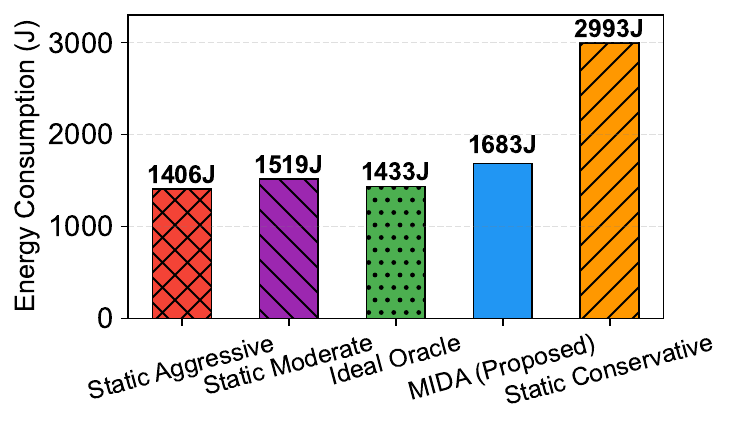}
        \caption{Energy consumption based on different timer setting schemes in PSM.}
        \label{psm3}
    \end{minipage}\hspace{0.02\textwidth}
    \begin{minipage}[t]{0.3\linewidth}
        \centering
        \includegraphics[width=1.05\linewidth,height=6cm,keepaspectratio]{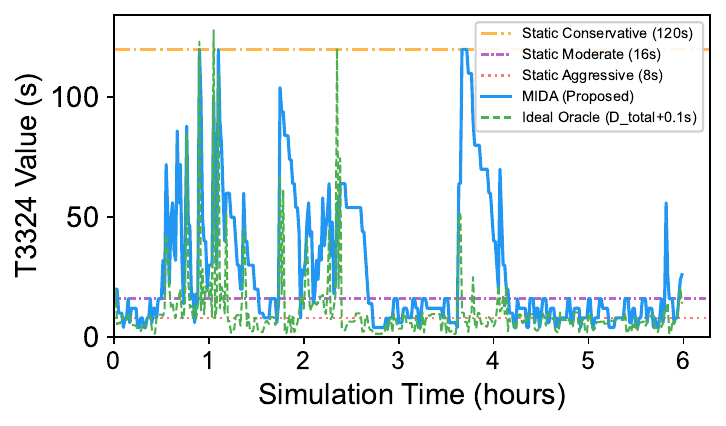}
        \caption{Energy-DDR Pareto frontier based on different timer setting schemes in PSM.}
        \label{psm4}
    \end{minipage}\hspace{0.02\textwidth}
    \begin{minipage}[t]{0.23\linewidth}
        \centering

    \end{minipage}
    \vspace{0.2cm}
\end{figure*}

As shown in Fig.~\ref{psm2} and Fig.~\ref{psm3}, MIDA achieves 85.2\% Downlink Delivery Ratio (DDR) with 1683 J average energy. It saves 43.8\% energy versus the static conservative baseline of 120 s which achieves 99.3\% DDR at the cost of 2993 J energy consumption. Meanwhile, it outperforms the static aggressive baseline of 8 s (20.5\% DDR, 1406 J) by 65\% while only bringing 20\% additional energy cost. Compared to the static moderate baseline of 16 s (70.1\% DDR, 1519 J), MIDA improves DDR by 15\% with a little 10.8\% energy overhead. These results demonstrate that feedback-driven PSM timer adaptation design outperforms the fixed PSM timer which can not adopt to the dynamic traffic pattern and thus lacks knowledge of the current delay regime. In contrast, MIDA adopts an adaptive PSM timer and aligns well with the traffic pattern.

As described in Fig.~\ref{psm4}, the ideal oracle scheme achieves 100\% DDR at the energy cost of 1433 J, establishing an energy-DDR Pareto frontier that MIDA approaches within 17\% in energy and 15\% in DDR. These results demonstrate that a simple, stateless, feedback-driven control law can closely track an omniscient oracle, outperforming all fixed-threshold baselines on the DDR–energy trade-off, and validating MIDA as a practical zero-configuration solution for NB-IoT PSM tuning in high-latency satellite deployments.

\begin{figure*}[htbp]
    \centering
    \begin{minipage}[t]{0.3\textwidth}
        \centering
        \includegraphics[width=1.05\linewidth,height=5cm,keepaspectratio]{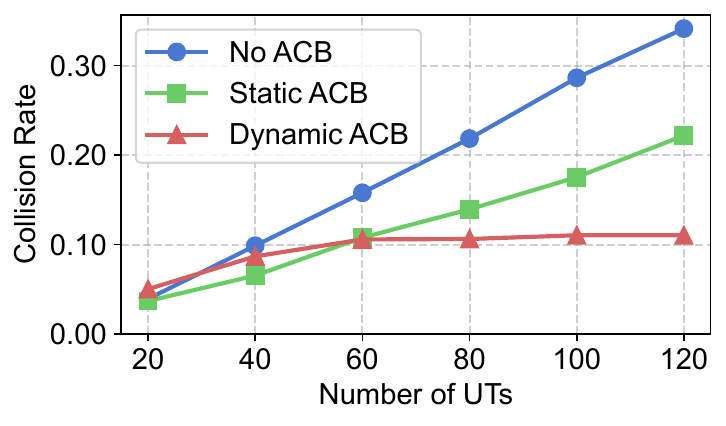}
        \caption{The collision rate based on different access schemes.}
        \label{ACB}
    \end{minipage}\hspace{0.02\textwidth}
    \begin{minipage}[t]{0.3\linewidth}
        \centering
        \includegraphics[width=1.05\linewidth,height=5cm,keepaspectratio]{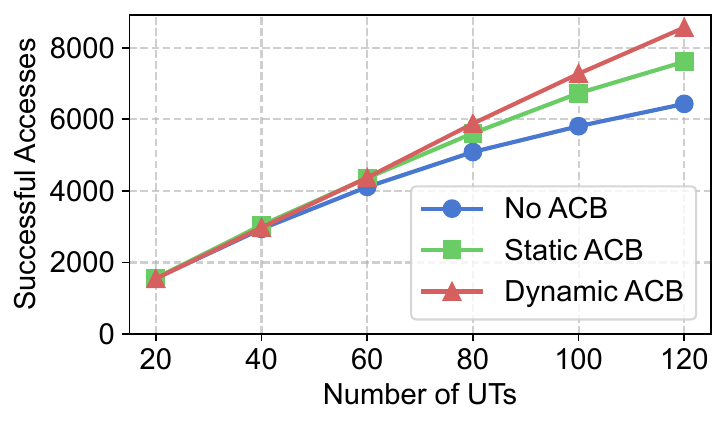}
        \caption{The number of success access based on different access schemes.}
        \label{ACB1}
    \end{minipage}\hspace{0.02\textwidth}
    \begin{minipage}[t]{0.3\linewidth}
        \centering
        \includegraphics[width=1.05\linewidth,height=5cm,keepaspectratio]{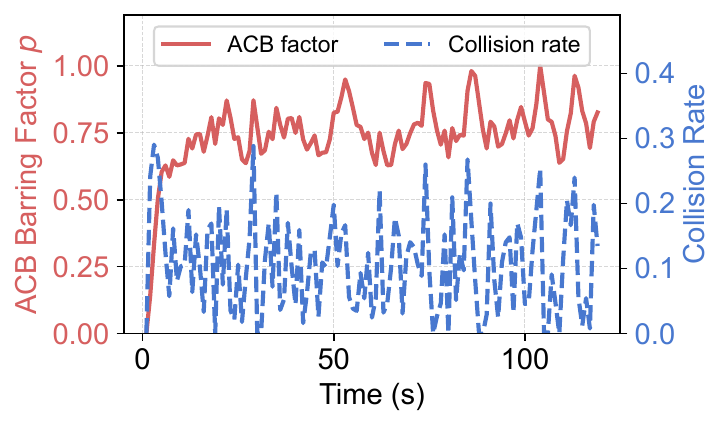}
        \caption{The barrier factor adjustment over time.}
        \label{ACB2}
    \end{minipage}\hspace{0.02\textwidth}
    \begin{minipage}[t]{0.23\linewidth}
        \centering

    \end{minipage}
    \vspace{-\baselineskip}
\end{figure*}

To further mitigate the access competition, a simple yet effective Access Class Barring (ACB) aware access mechanism is designed, the intuition behind which is to leverage the parameter--ACB that is broadcast by the satellite to determine whether the ground terminal can be accessed. Actually, this ACB-aware mechanism is to dynamically adjust the barring factor according to the per-second collision rate sensed at the satellite side, which can be formulated as below.
\begin{equation}
   p=p+Kp\times (measured\_rate-target\_rate), 
\end{equation}

\noindent where $Kp$ denotes the proportional gain. $measured\_rate$ and $target\_rate$ represent the measured and targeted collision rate, respectively. In this equation, large deviation incurs fast response, and small deviation results in smooth convergence. When the ground terminal passes probability detection, it can be allowed to be accessed, thereby lowering down the collision probability. Next, we have conducted simulations to verify the efficiency of this ACB-aware access mechanism. As shown in Fig.~\ref{ACB}, the collision rate based on dynamic ACB maintains a low level of 10\% when the number of ground terminals exceeds 100, while this metric based on static ACB increases from 13\% to 23\%. Even worse, it increases from 23\% to 34\% simultaneously if only ALOHA is adopted. It verifies the high efficiency of the proposed access mechanism.

Simultaneously, the ACB-aware access scheme brings a higher successful access probability compared to the available ALOHA protocol, as verified in Fig.~\ref{ACB1}. To investigate in detail, we have also measured the key indicator--barrier factor in the ACB-aware access mechanism. Fig.~\ref{ACB2} describes the adjustment of barrier factor with the increase of ground terminals. Especially, when the number of ground terminals increases from 20 to 100, the barrier factor increases from 0.10 to 0.68, in order to intercept 68\% of access requests and reduce the collision rate from 34.3\% to 11.2\% with a decrease of 67\%.

\section{Discussion and Limitations}

\textbf{Limited concurrency degree.} We have deployed ten ground terminals to mimic the concurrent transmission scenario, and then investigate the end-to-end delay under different concurrency levels. One may wonder there should exist more ground terminals (e.g., hundreds or thousands) attempting to connect with the GEO satellite in the actual case. However, we have acquired the delay distribution of different ground terminals, which demonstrates the E2E delay distribution trend coupled with the increase of ground terminals. Naturally, the delay distribution will be widened with the increase of concurrency degree due to more intense access competition.

\textbf{Incomprehensive measurements.} As described above, this paper focused on measuring two critical metrics--E2E delay and power consumption in NB-IoT NTN, while ignoring other perspectives such as link reliability under different weather conditions or using different kinds of antennas. We leave these kinds of measurements for future work.  


\section{Related Work}
Research work on integrating narrowband cellular IoT with non-terrestrial networks (NB-NTN) sits at the intersection of 3GPP standardization, analytic modeling, prototype validation, and emerging commercial rollouts \cite{medina20253gpp,el2023introduction}. 3GPP has driven the protocol and system architecture for NTNs across multiple releases (studies/specs for NR-NTN and the NB-IoT/eMTC extensions), and its public overview and technical reports remain the primary reference for supported deployment scenarios, channel models, and the set of NTN design options (GEO/LEO/HAP, waveform adaptations, and regulatory considerations) \cite{xia20245g,liberg2021narrowband,medina20243gpp}. 

A growing literature studies the performance and protocol implications of NB-IoT over satellites \cite{trinc2025comparative,amatetti2022nb,beale2021iot}. Several survey and design papers explain how standard NB-IoT must be adapted for high delay, large Doppler, and long visibility windows, and produce analytic models for random access, uplink scheduling, and link budget tradeoffs for LEO/GEO deployments \cite{amatetti2024novel,harounabadi2023toward,sciddurlo2021looking}. In particular, careful work on RACH/preamble detection and uplink synchronization shows how NB-IoT random-access procedures can be modified to improve detection probability and support the wide timing/frequency uncertainty introduced by satellite channels. These analytic and simulation studies provide useful baselines for expected RACH success, coverage extension, and power/latency tradeoffs \cite{li2025review,testi2025nb}. 

Complementing theory, several prototype and testbed efforts (including academic CubeSat experiments and dedicated in-orbit test missions) report real implementation issues: RF front-end constraints, antenna and pointing tradeoffs, Doppler compensation, gateway to satellite integration, and the operational burden of certification and testing \cite{amatetti2022nb,kim2022performance}. Recent engineering testbeds and mission descriptions (ground–in-the-loop validation and small LEO missions) demonstrate both feasibility and a number of practical “gotchas” that are rarely visible in analytic work (e.g., intermittent contacts, payload processing limits, and test instrumentation constraints). These experimental papers and industry roadmaps are critical precedents to any deployment-oriented experience report \cite{hou2025sdr,azari2022evolution,therrien2022large}. 

Despite rich standardization, modeling, and prototype activity, what remains scarce in the literature is a large-scale, reproducible experience account that systematically documents long-term operational telemetry, the engineering decisions required to integrate off-the-shelf NB-IoT hardware with satellite payloads, and the empirical divergence between model predictions and measured system behavior.

\section{Conclusion}
In this paper, we have conducted measurements and investigated the initial performance for IoT-NTN networks over GEO. We propose several simple yet effective designs to overcome the inefficiencies obtained from the measurement and analysis, which lay a good foundation for the future implementation and deployment of IoT-NTN over GEO.

\bibliographystyle{plain}
\bibliography{reference}

\end{document}